\documentclass[prb,twocolumn,amsmath,showpacs]{revtex4}
\input {epsf.sty}
\usepackage{epsfig}
\usepackage{amsmath,amsthm,amssymb,latexsym,amsfonts}

\begin{document}

\title{Stability of the A-like Phase of 
Superfluid $^3$He\\ in  Aerogel with Globally 
Anisotropic
Scattering}

\author{J.P. Davis, J. Pollanen, B. Reddy, K.R. 
Shirer, H. Choi and W.P. Halperin}
\affiliation{Department of Physics and Astronomy,
\\ Northwestern University, Evanston, IL 60208, USA}

\date{Version \today}


\begin{abstract} It has been suggested that 
anisotropic quasiparticle  scattering will 
stabilize
anisotropic phases of superfluid $^3$He contained 
within highly  porous silica aerogel. For example,
global anisotropy introduced via uniaxial 
compression of aerogel might stabilize the axial 
state,
which is called the A-phase in bulk  superfluid 
$^3$He.  Here we present measurements of the phase
diagram of  superfluid $^3$He in a 98\%  porous 
silica aerogel using transverse acoustic impedance
methods.  We  show that uniaxial  compression of 
the aerogel by 17\% does not stabilize an axial
phase.
\end{abstract}

\maketitle

When disorder is introduced into superfluid 
$^3$He by way of  high porosity silica aerogel a
metastable A-like phase appears on cooling 
\cite{Bar00, Ger01, Ger02, Naz04}.  This phase is 
thought
to be like the A-phase in bulk  superfluid 
$^3$He, known to be the axial $p$-wave state. At
sufficiently low temperatures this  metastable 
phase undergoes a transition to an isotropic
superfluid phase similar to the isotropic state 
observed in bulk $^3$He, the B-phase. However,  a
distinct transition from the B-like phase to the 
A-like  phase in aerogel is not seen upon warming.
Tracking experiments
\cite{Ger02, Naz04, Bau04,Vic05} have shown that coexistence of
   A-like and B-like phases occurs in a narrow window of temperature,
$\approx 20$ - $50~\mu$K, near the 
normal-to-superfluid transition temperature in 
aerogel,
$T_{ca}$.  This is contrary to the expectation 
that the B-phase  should be stable at all 
pressures
and temperatures if the disorder introduced is 
homogenous  and the scattering is isotropic
\cite{Thu98}.  On the other hand, it has been 
predicted that  scattering anisotropy from the 
strands
of aerogel might destabilize the B-like phase in 
favor of  the A-like phase \cite{Thu98}.

Pursuing this idea, Vicente \emph{et al.} 
\cite{Vic05} suggested that  the introduction of
\emph{global} anisotropy into aerogel, for 
example by uniaxial strain, might  increase the 
stability
of the A-like phase.  Recent calculations 
\cite{Aoy06} have shown that uniaxial  anisotropy 
(achieved
for example by compression along one axis) should 
stabilize the axial  state, whereas radial
anisotropy (radially compressed or radially 
reduced by preferential shrinkage  during growth) 
might
stabilize the polar state.  Our previous results
\cite{Dav06} for $^3$He in aerogel with 
preferential radial shrinkage  suggest a phase 
with increased
stability, but the aerogel was not rigidly 
adhered to the transducer  surface so there is 
some
question as to whether or not this was an effect 
intrinsic to superfluid $^3$He  in aerogel. In 
this
paper we present our measurements of the phase 
diagram for superfluid $^3$He in a  sample of 98\%
porosity silica aerogel grown directly on the 
surface of a transducer and then subjected to 
uniaxial
strain of 17\%.

	We used transverse acoustic impedance 
measurements \cite{Ger01,  Ger02, Dav06} at the 
third harmonic
(17.6 MHz) of an \emph{AC}-cut quartz 
piezoelectric transducer, 0.84  cm in diameter. 
The impedance
was measured using a frequency modulated 
\emph{RF}-bridge, described  elsewhere 
\cite{Ham89}.  It has
been shown
\cite{Ger01, Ger02} that for aerogel grown 
directly onto the  transducer surface, the 
measured
impedance is sensitive to all phase transitions 
through coupling of the shear  transducer to the
superfluid and is coincident with transitions in 
the interior of the aerogel.  We grew our aerogel
sample in the  open space between two parallel transducers separated by 
two spacer wires, 0.0305~cm
diameter, held under tension from a stainless 
steel spring, Fig.~1.  Two additional spacer 
wires of
smaller diameter, 0.0254~cm,  were placed along 
side and between the  larger ones before aerogel 
was
grown to fill the entire assembly.

\begin{figure}[b]
      \centerline{\includegraphics[width=2.8in]{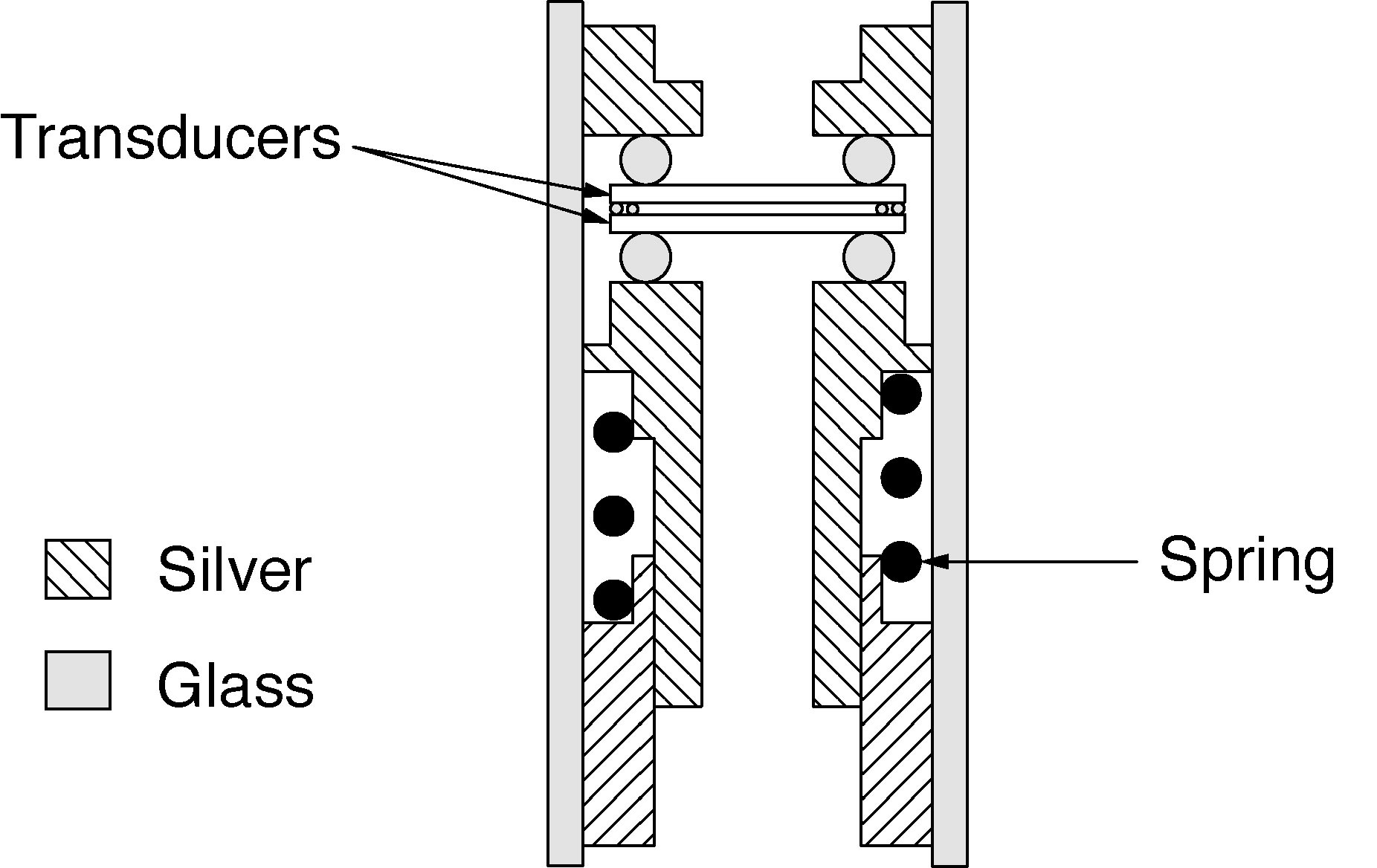}}
\caption {\label{fig1} Assembly diagram for the 
support structure  that allows the aerogel to be
grown directly between two quartz transducers and 
then compressed  {\it in situ}.  After aerogel
growth and compression a glass sleeve on the 
outside of the assembly was  epoxied in place.}
\end{figure}

	The aerogel was synthesized at 
Northwestern University via a  one-step sol-gel 
process followed by
supercritical drying \cite{Pol07}.  The density 
was controlled by the  ratio of the reactants 
during
the synthesis and was measured after drying to be 
97.8\% porous.  After drying, the excess aerogel
was removed, leaving only the aerogel between the 
two parallel  transducers such that their outer
surfaces could be exposed to bulk $^3$He.  Next, 
the 0.0305~cm diameter spacers were removed,
maintaining tension with the spring, such that 
the aerogel was compressed to 0.0254~cm, giving 
17\%
uniaxial strain.  This amount of compression was 
shown by Pollanen {\it et al.} \cite{Pol07} to
result in global anisotropy on the length scale 
of the correlation length of aerogel,
$\approx 20$ nm, causing minimal plastic 
deformation, as measured by small angle x-ray 
scattering
(SAXS).  Additionally, Pollanen \emph{et al.} 
have used optical birefringence measurements  to
demonstrate that this method of imposing strain 
transmits anisotropy uniformly  from macroscopic
length scales to the microscopic scale probed by 
SAXS. Samples of the aerogel  removed from the
assembly region, adjacent to the acoustic sample, 
were also characterized using optical
birefringence techniques
\cite{Pol07} to ensure that, before compression, 
our aerogel sample  was isotropic and homogeneous.

\begin{figure}[t]
      \centerline{\includegraphics[width=3.4in]{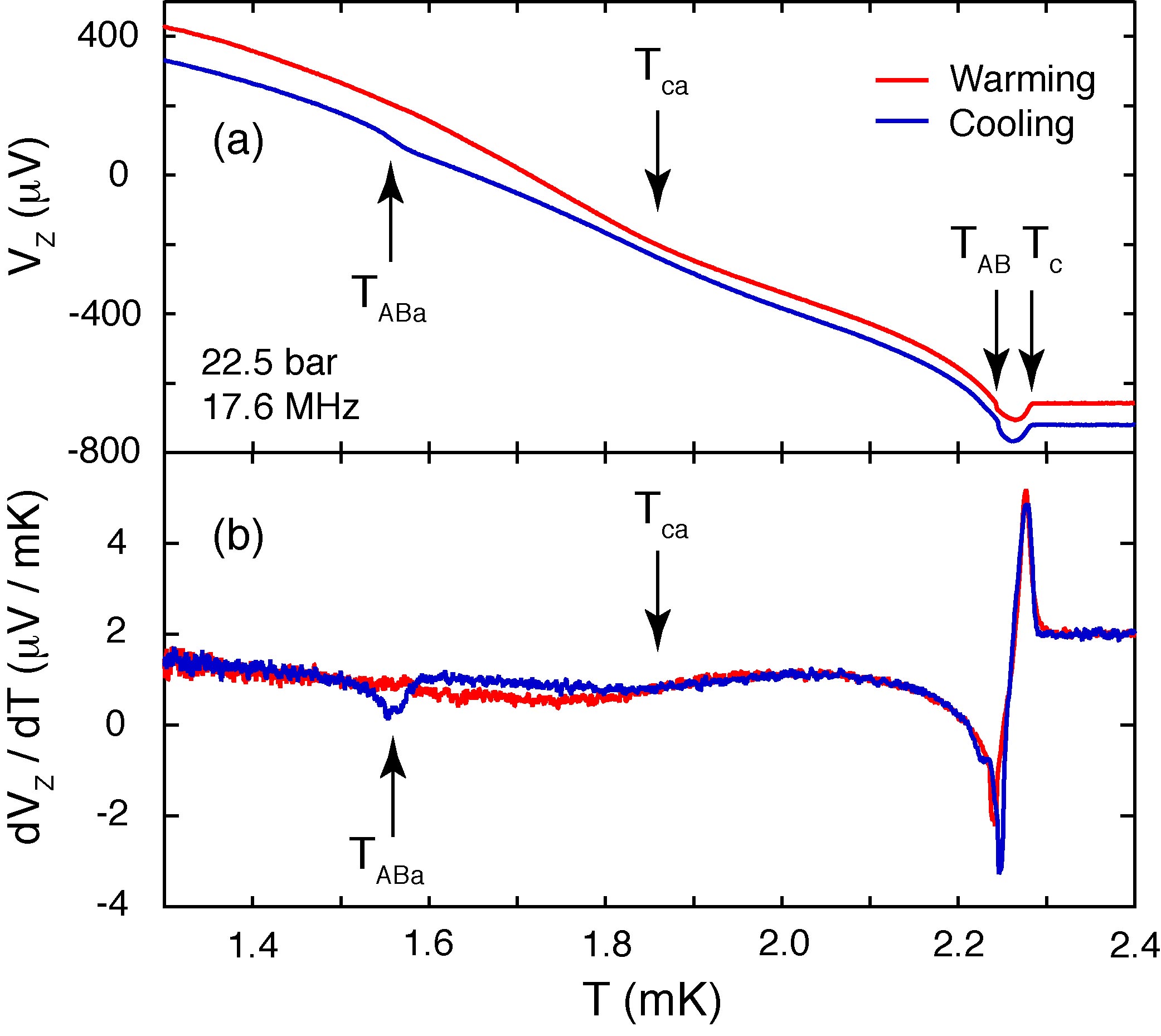}}
\caption {\label{fig2} (Color online) Transverse 
acoustic impedance  measurements, $V_{z}$, as a
function of temperature at 22.5 bar.  (a) The bulk superfluid transitions are
$T_{c}$ and $T_{AB}$ and are very distinct. The 
aerogel phase transitions, $T_{ca}$ and $T_{ABa}$,
are  more spread out, but are easier to identify 
in the derivative of acoustic response with 
respect
to  temperature, in (b).  Note that the warming 
and cooling trace do not match up in the 
temperature
interval  between $T_{ca}$ and $T_{ABa}$. On 
cooling, this region corresponds to the 
supercooled
A-like phase,  whereas on warming, this corresponds to the B-like phase.}
\end{figure}

\begin{figure}[t]
   \centerline{\includegraphics[width=3.0in]{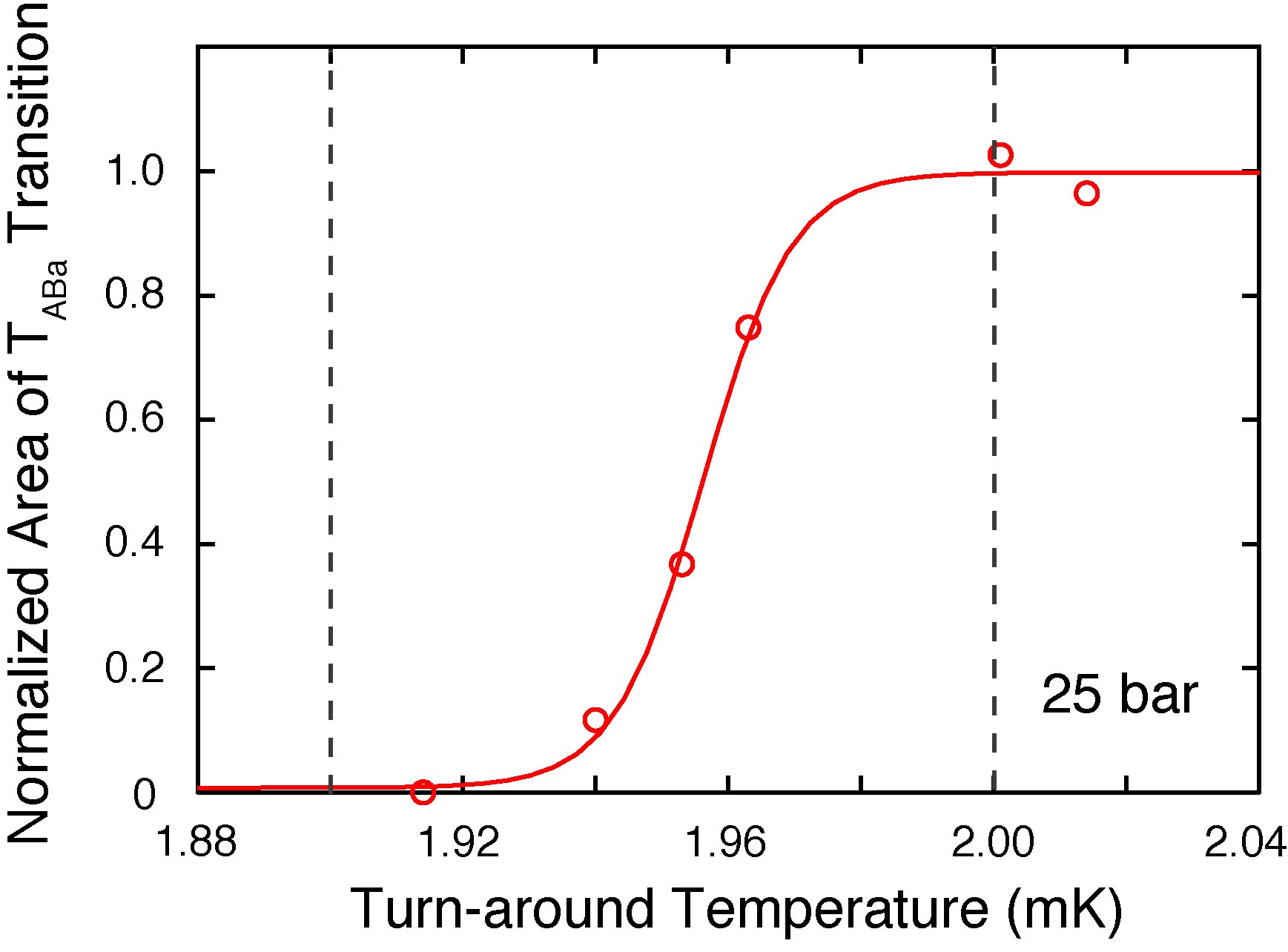}}
\caption {\label{fig3} (Color online) The 
normalized area of the  dip in the $dV_{Z}/dT$ 
versus $T$
trace, such as shown in Fig.~2b, at 25 bar.  The 
transition width is  approximately $40~\mu$K,
similar to that of uncompressed aerogel 
\cite{Ger02, Vic05}. The dashed vertical lines 
are our estimate of the precision with which we can 
independently identify $T_{ca}$ which overlaps the
coexistence region of A-like and B-like phases.}
\end{figure}

	The aerogel sample and experimental 
assembly were cooled in  liquid $^3$He using a 
dilution refrigerator, followed by adiabatic nuclear demagnetization
\cite{Ham89}.  A SQUID-based paramagnetic salt 
(LCMN) thermometer was used \cite{Ham89}, 
calibrated
from the  Greywall temperature scale
\cite{Gre86} using bulk superfluid $^3$He 
transitions that were  easily identified in the 
acoustic
response, $V_{z}$, Fig.~2a.  We
determined the temperature of the aerogel phase transitions by taking  the derivative 
of the acoustic
response with respect to the temperature, 
Fig.~2b.  The transition  temperature from
normal-to-superfluid in aerogel is best indicated 
by the point of separation of the  warming and
cooling traces as shown in Fig.~2b.  Transitions 
from the A-like phase to the B-like phase are seen
upon cooling, appearing as a dip in the 
derivative trace.  No such transition is seen on 
warming.
Similar signatures of these phase transitions 
have been reported previously for isotropic 
aerogel \cite{Ger01, Ger02}.

	Gervais \emph{et  al.} \cite{Ger02} and Vicente \emph{et al.} \cite{Vic05} 
performed tracking
experiments by warming up close to, but not 
through, the aerogel  superfluid transition 
temperature,
$T_{ca}$.  After stopping at a `turn-around' 
temperature the samples  were then cooled again 
to look
for an A-like to B-like transition.  In this way 
it is possible to find the warming  transition and
how close the turn-around temperature must be to the critical temperature,
$T_{ca}$, to observe it.  The magnitude of the 
impedance change is a measure of the amount of
superfluid undergoing the A-like to B-like 
transition.  We performed these  tracking 
experiments at
25~bar in order to determine the window of 
coexistence of A-like and  B-like phases in 
uniaxially
compressed aerogel.  We integrated the area of 
the dip in the derivative of  the acoustic 
response
with temperature and plot this as a  function of 
the `turn-around' temperature in Fig.~3 at 25~bar.
The coexistence region is
$\approx 40~\mu$K which, to within our precision, 
is within  $50~\mu$K of $T_{ca}$ similar to that
reported earlier\cite{Ger02,Vic05} for nominally isotropic aerogel.

\begin{figure}[t]
      \centerline{\includegraphics[width=3.4in]{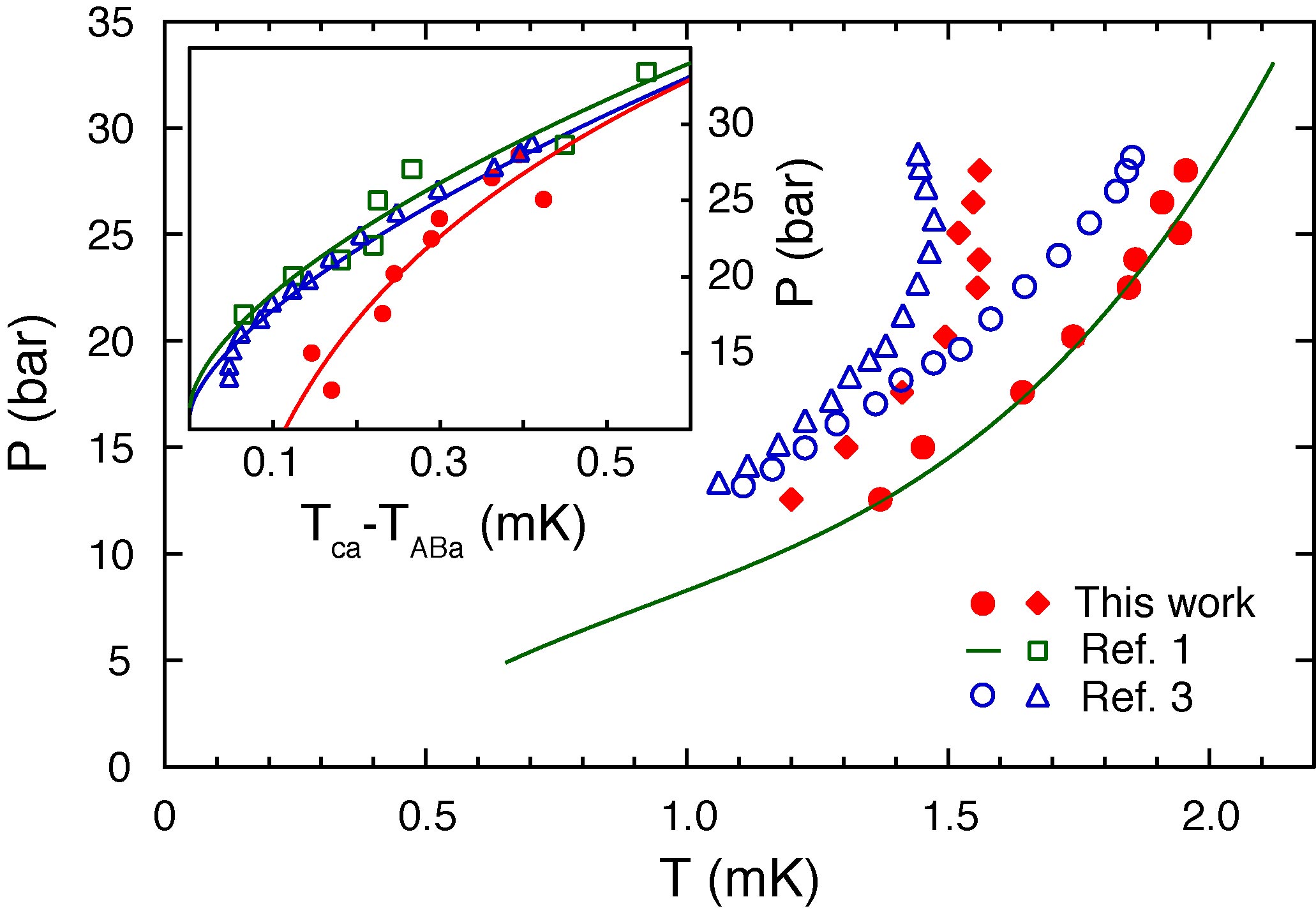}}
\caption {\label{fig4} (Color online) Pressure 
versus temperature  phase diagram for 17\% 
uniaxially
compressed aerogel (filled red symbols).  Both 
the aerogel  superfluid transition temperature 
and the
extent of supercooling of the A-like phase are 
similar to previous results with uncompressed
aerogel: Gervais \emph{et al.} \cite{Ger01} (green curve and squares) and Nazaretski \emph{et al.} \cite{Naz04} (open blue  symbols). Inset: 
The difference between the  temperatures for the
aerogel normal-to-superfluid transition
$T_{ca}$ and the supercooled A-like to B-like 
$T_{ABa}$ transitions as  a function of pressure. 
Note
that there is more supercooling in the compressed 
aerogel at low pressures  than in previous work on
uncompressed samples. Curves in the inset are guides to the eye.}
\end{figure}

	In Fig.~4, we show the superfluid 
transitions, $T_{ca}$, as well as the amount of 
supercooling in
our uniaxially compressed aerogel compared to 
that of Gervais \emph{et al.} \cite{Ger01, Ger02} 
and
Nazaretski
\emph{et al.} \cite{Naz04}.  The similarity is 
striking given the significant amount of global
anisotropy in our sample.  The only apparent 
difference between our results on axially 
compressed
aerogel and previous work is the increase in the 
supercooling of the A-like phase at pressures 
below
20~bar.  This does not bear directly on the 
stability of the A-like phase, but suggests that 
the
mechanism  for nucleation of the B-phase is 
suppressed at lower pressures for uniaxially 
anisotropic
aerogel.  We have also found that the signature 
of the A-like to B-like transition  becomes 
smaller
as the pressure is decreased until it becomes 
difficult to measure below 12~bar. Although we 
find
that uniaxial compression  of the aerogel does 
not enhance phase stability, nonetheless we note 
that
there are recent reports that the orientation of 
the superfluid order parameter can be influenced 
by
anisotropy \cite{Kun07, Elb07, Dim07a, Dim07b}.

	In summary, we find that the introduction 
of global  anisotropy from uniaxial compression 
of 17\%
does not stabilize the A-like phase of superfluid 
$^3$He in aerogel, in  contrast to various
suggestions
\cite{Vic05,Aoy06}.  The region of coexistence of the 
A-like and  B-like phases is approximately
$40~\mu$K and indistinguishably close to the 
normal-to-superfluid transition, nearly the same 
as that
measured previously in  uncompressed aerogel 
\cite{Ger02,Vic05}. Consequently, it appears that
uniaxial strain does not stabilize an  A-like 
phase, or for that matter any phase, in aerogel. 
The pressure versus temperature phase diagram  is 
remarkably similar to uncompressed aerogel, 
except for increased supercooling at low  pressures in the range, 12 - 20~bar.

	We acknowledge support from the National 
Science Foundation,  DMR-0703656 and thank W.J. Gannon for useful discussions.


\begin{thebibliography}{xxx}

\bibitem{Bar00} B.I. Barker, Y. Lee, L. Polukhina, D.D. Osheroff, L.W. Hrubesh and J.F. Poco, Phys. Rev. Lett. {\bf 85}, 2148 (2000).

\bibitem{Ger01} G. Gervais, T.M. Haard, R. 
Nomura, N. Mulders and W.P. Halperin, Phys. Rev. 
Lett.
\textbf{87}, 035701 (2001).

\bibitem{Ger02} G. Gervais, K. Yawata, N. Mulders 
and W.P. Halperin, Phys. Rev. B \textbf{66}, 
054528
(2002).

\bibitem{Naz04} E. Nazaretski, N. Mulders and 
J.M. Parpia, JETP Lett. \textbf{79}, 383 (2004).

\bibitem{Bau04} J.E. Baumgardner, Y. Lee, D.D. 
Osheroff, L.W. Hrubesh and J.F. Poco, Phys. Rev. 
Lett.
{\bf 93}, 055301 (2004).

\bibitem{Vic05} C.L. Vicente, H.C. Choi, J.S. 
Xia, W.P. Halperin, N.  Mulders and Y. Lee, Phys. 
Rev.
B \textbf{72}, 094519 {2005}

\bibitem{Thu98} E.V. Thuneberg, S.K. Yip, M. 
Fogelstr\"om and J.A. Sauls, Phys. Rev. Lett.
\textbf{80}, 2861 (1998).

\bibitem{Aoy06} K. Aoyama and R. Ikeda. Phys. Rev B \textbf{73}, 060504 (2006).

\bibitem{Dav06} J.P. Davis, H. Choi, J. Pollanen 
and W.P. Halperin,  AIP Conf. Proc. \textbf{850},
239 (2006).

\bibitem{Kun07} T. Kunimatsu, T. Sato, K. 
Izumina, A. Matsubara, Y. Sasaki, M. Kubota, O. 
Ishikawa,
Y. Mizusaki and Y.M. Bunkov, JETP Lett. \textbf{86}, 216 (2007).

\bibitem{Elb07} J. Elbs, Y.M. Bunkov, E. Collin, 
H. Godfrin and G.E. Volovik (2007).
[http://arxiv.org/abs/0707.3544]

\bibitem{Dim07a} V.V. Dmitriev, D.A. Krasnikhin, 
N. Mulders, V.V. Zavjalov and D.E. Zmeev, JETP 
Lett.
\textbf{86}, 594 (2008).

\bibitem{Dim07b} V.V. Dmitriev, D.A. Krasnikhin, 
N. Mulders and D.E. Zmeev, J. of Low Temp. Phys.
\textbf{150}, 493 (2008).

\bibitem{Ham89} P.J. Hamot, H.H. Hensley and W.P. 
Halperin, J. of Low Temp. Phys \textbf{77}, 429
(1989).

\bibitem{Pol07} J. Pollanen, K. Shirer, S. 
Blinstein, J.P. Davis, H. Choi, T.M. Lippman, 
L.B. Lurio
and W.P. Halperin (2007). [http://arxiv.org/abs/0711.3495]

\bibitem{Gre86} D.S. Greywall, Phys. Rev. B \textbf{33}, 7520 (1986).


\end{thebibliography}
\end{document}